\documentclass{aa}

\usepackage{txfonts}

\usepackage{amsfonts,amssymb,amsbsy,amsmath}
\usepackage{bigints}
\usepackage{booktabs}
\usepackage{footnote}
\usepackage{graphics}                    
\usepackage{graphicx}
\usepackage{hyperref} 
\usepackage{mathrsfs}
\usepackage{mathtools}
\usepackage{natbib}
\usepackage{nicefrac}
\usepackage{rotating}
\usepackage{slashbox}
\usepackage{subfig}
\usepackage[theorems]{tcolorbox}
\usepackage{upgreek}
\usepackage{color}
\usepackage{xifthen}
\usepackage[normalem]{ulem} 

\DeclareRobustCommand{\orderof}{\ensuremath{\mathcal{O}}} 

\catcode`_=\active
\newcommand_[1]{\ensuremath{\sb{\mathrm{#1}}}}

\newcommand{\newpar}{{\\}}

\newcommand{\diff}{{\operatorname{d}}}

\newcommand{\bs}{\boldsymbol}

\newcommand{\numChar}{{n}} 

\newcommand{\paraChar}{{p}}

\newcommand{\episChar}{{n}}

\newcommand{\npe}[1][]{{ \numChar_{ \ifthenelse{\isempty{#1}}{\paraChar\episChar}{{\paraChar\episChar,#1}} } }} 

\newcommand{\epis}{{nois}}

\def\gtrsim{\mathrel{\hbox{\rlap{\hbox{\lower4pt\hbox{$\sim$}}}\hbox{$>$}}}}
\def\lessim{\mathrel{\hbox{\rlap{\hbox{\lower4pt\hbox{$\sim$}}}\hbox{$<$}}}}

\newcommand{\rmz}{{\rm z}}

\newcommand{\emodel}[1][]{{ M_{ \ifthenelse{\isempty{#1}}{\epis}{{\epis,#1}} } }}
\newcommand{\emodelz}[1][]{{ M_{ \rmz\ifthenelse{\isempty{#1}}{\epis}{{\epis,#1}} } }}
\newcommand{\emodellgrb}[1][]{{ M^\lgrb_{ \ifthenelse{\isempty{#1}}{\epis}{{\epis,#1}} } }}

\newcommand{\lgrb}{{\rm g}}

\newcommand{\param}{{\bs{\theta}}}

\newcommand{\eparam}[1][]{{ \param_{ \ifthenelse{\isempty{#1}}{\epis}{{\epis,#1}} } }}
\newcommand{\eparamz}[1][]{{ \bs\param^\rmz_{ \ifthenelse{\isempty{#1}}{\epis^\rmz}{{\epis,#1}} } }}
\newcommand{\eparamlgrb}[1][]{{ \bs\param^\lgrb_{ \ifthenelse{\isempty{#1}}{\epis^\lgrb}{{\epis,#1}} } }}

\newcommand{\truth}{{\bs{R}}}
\newcommand{\possible}{{*}}
\newcommand{\truthset}{{\mathcal{R}}}

\newcommand{\truthsubset}[1][]{{ \truthset_{ \ifthenelse{\isempty{#1}}{\truth}{{\truth_{#1}}} } }}
\newcommand{\ptruthsubset}[1][]{{ \truthset_{ \ifthenelse{\isempty{#1}}{\truth}{{\truth_{#1}}} }^\possible }}

\newcommand{\xx}[1][]{{ \ifthenelse{\isempty{#1}}{\textcolor{red}{XXX}}{\textcolor{red}{~(XXX {#1} XXX)~}} }}

\newcommand{\dur}{{T_{90}}}

\defcitealias{hopkins2006normalization}{H06}
\defcitealias{li2008star}{L08}
\defcitealias{butler2010cosmic}{B10}
\defcitealias{madau2017radiation}{M17}
\defcitealias{fermi2018gamma}{F18}

\defcitealias{yu2015unexpectedly}{Y15}
\defcitealias{petrosian2015cosmological}{P15}
\defcitealias{pescalli2016rate}{P16}
\defcitealias{tsvetkova2017konus}{T17}
\defcitealias{lloyd2019cosmological}{L19}

\begin{document}
	\title{An alternative statistical interpretation for the apparent plateaus in the duration distributions of GRBs}

    \author{
        Joshua Alexander Osborne\inst{1}\fnmsep\thanks{joshua.osborne@uta.edu}
        \and
        Christopher Michael Bryant\inst{1}
        \and
        Fatemeh Bagheri\inst{1}\fnmsep\thanks{fatemeh.bagheri@uta.edu (NSF-ASCEND Postdoctoral Fellow)}
        \and
        Amir Shahmorad\inst{1}\inst{2}\fnmsep\thanks{shahmoradi@utexas.edu (corresponding author)}
    }

    \institute{
        Department of Physics,
        The University of Texas,
        Arlington, TX 76010, USA
        \and
        Division of Data Science 
        The University of Texas,
        Arlington, TX 76010, USA
    }

    \date{Received \today; accepted pending}

    \abstract
    {
        The existence of a plateau in the short-duration tail of the observed distribution of cosmological Long-soft Gamma Ray Bursts (LGRBs) has been argued as the first direct evidence of Collapsars. A similar plateau in the short-duration tail of the observed duration distribution of Short-hard Gamma Ray Bursts (SGRBs) has been suggested as evidence of compact binary mergers.
    }
    {
        We present an equally plausible alternative interpretation for this evidence, which is purely statistical.
    }
    {
        Specifically, we show that the observed plateau in the short-duration tail of the duration distribution of LGRBs can naturally occur in the statistical distributions of strictly positive physical quantities, exacerbated by the effects of mixing with the duration distribution of SGRBs, observational selection effects, and data aggregation (e.g., binning) methodologies. The observed plateau in the short-duration tail of the observed distributions of SGRBs can similarly result from a combination of sample incompleteness and inhomogeneous binning of data. The observed plateau in the short-duration tail of the observed distributions of SGRBs can similarly result from a combination of sample incompleteness and inhomogeneous binning of data. We further confirm the impact of these factors on the observation of a plateau in the duration distributions of Gamma-Ray Bursts (GRBs) through extensive numerical Monte Carlo simulations.
    }
    {
        Our presented analysis corroborates and strengthens a purely statistical and sample-incompleteness interpretation of the observed plateau in the duration distribution of LGRBs and SGRBs without invoking the physics of Collapsars and jet-propagation through the stellar envelope.
    }
    {}


    \maketitle

\section{Introduction}
\label{sec:intro}

	Gamma-ray bursts (GRBs) are primarily broken into two categories: Long-duration GRBs (LGRBs), with $T_{90} > 2$ s, and Short-duration GRBs (SGRBs), with $T_{90} < 2$ s \citep[e.g.,][]{kouveliotou1993identification}. $\dur$ is the time it takes to receive 90\% burst fluence, starting when 5\% of the fluence has been observed \citep[e.g.,][]{Poolakkil2021}. 
	Significant observational evidence over the past two decades connects LGRBs to the death and the iron core-collapse of supermassive stars \citep{fruchter2006long, woosley2006supernova}. Specifically, the leading theory points toward Wolf–Rayet stars, stars with masses between $10M_\odot-50M_\odot$ \citep{massey1981masses}, as being the most likely progenitors of LGRBs due to the mass loss incurred due to stellar wind, allowing for the jets to break out of the stellar envelope. This Collapsar model of LGRBs was first supported by observational evidence through the association of half a dozen GRBs with spectroscopically confirmed broad-line Ic supernovae (SNe) as well as photometric evidence of underlying SNe in about two dozen more \citep{galama1998unusual, stanek2003spectroscopic, hjorth2003very, woosley2006progenitor, hjorth2011gamma}. In addition, there is indirect evidence for the connection of LGRBs with massive stars from the identification of LGRB host galaxies as intensively star-forming galaxies \citep{bloom2002detection, le2003hosts, christensen2004uv, fruchter2006long}. The LGRBs are localized in the most active star-forming regions within those galaxies, increasing the probability that they originate from the death of supermassive stars. Physically, according to the Collapsar model, as the core collapses in a supernova ({\bf SN}) explosion, a bipolar jet is launched from the center of the star that has to drill through the stellar envelope and break out to produce an observed GRB \citep[e.g.,]{macfadyen1999collapsars, macfadyen2001supernovae}.


	
	While the evidence for LGRBs-supernovae association \citep{bloom2001prompt} provided a strong support for the Collapsar model, the first strong support for the jet-ejecta interaction came from the off-axis afterglow observations of GRB 170817A \citep[e.g.,][]{abbott2017gw170817, goldstein2017ordinary, savchenko2017integralDetection}. In their paper, \citet{bromberg2012observational} develop a model of LGRBs that provides the first prompt-emission observational imprint of this jet-envelope interaction, and thus, direct confirmation of the Collapsar model. A distinct signature in the duration distribution is detected: the appearance of a plateau toward short durations. This occurs at times much shorter than the typical breakout time of the jet, i.e., the time it takes for the jet to drill through the stellar envelope. This action dissipates energy, so the engine driving these jets must be in operation for at least the breakout time. If it is not, a GRB is not produced. The GRB is brief for breakout times that are very close to the engine time, and a characteristic plateau is seen in the duration distribution. The breakout time is set by the density and radius of the stellar envelope at radii $> 10^{10}$ cm. In comparison, the stellar core properties at radii $< 10^{8}$ cm determine the engine working time.
	\newpar
	
	Unlike LGRBs, SGRBs are theorized to result from the merger of either two neutron stars (NS-NS) or of a neutron star and a black hole (NS-BH) \citep{eichler1989nucleosynthesis, narayan2001accretion, metzger2012most, berger2014short, d2015short}. Recently, the first NS-NS merger was confirmed in an unprecedented joint gravitational and electromagnetic observation by Advanced Laser Interferometer Gravitational-Wave Observatory (LIGO), Advanced Virgo, INTErnational Gamma-Ray Astrophysics Laboratory (INTEGRAL) and the Fermi Gamma-ray Space Telescope's Gamma-ray Burst Monitor (GBM) \citep{abbott2017gw170817, goldstein2017ordinary, savchenko2017integral}. Aside from the finding of gravitational waves, this was also the first spectroscopically-confirmed event associated with a kilonova. Other recent studies using photometry have also found a kilonova event associated with GRB 211211A and GRB 230307A, both consisting of a hard to soft prompt emission \citep{troja2022nearby, levan2023jwst}. Compact binary mergers of these types are accompanied by significant dynamical mass ejection. Models range on how much mass is ejected depending on the GRB, from 0.03 - 0.08 \(\textup{M}_\odot\) (GRB 130603B) to 0.1 - 0.13 \(\textup{M}_\odot\) (GRB 060614) \citep{tanvir2013kilonova, berger2013r, yang2015possible}. The beamed nature of these events has been observed by inspecting breaks in optical and radio bands in GRB 990510 \citep{harrison1999optical}. Similar observations of breaks in SGRB afterglows have suggested the involvement of jets as well \citep{soderberg2006relativistic,fong2012jet}. Further supporting evidence for relativistic jets in GRBs is the observation of superluminal motion in GRBs by \citet{taylor2004angular, ghirlanda2019compact, mooley2018superluminal}, a phenomenon which is typically seen in the jets of blazars and quasars. Therefore, similar to the Collapsar model, the merger launches a relativistic jet that has to push through the expanding ejecta of significant mass \citep{murguia2014necessary, nagakura2014jet, kumar2015physics, duffell2015narrow, gottlieb2018cocoon}. According to \citet{moharana2017observational}, this again produces a plateau in the $\dur$ distribution, only at shorter durations than the one seen with LGRBs, reflecting the time it takes for prompt $\gamma$-ray emission to become observable after the jet reaches the ejecta's outer edge.
	
	
	In this manuscript, we argue that although the plateaus seen in the analysis of the $\dur$ distributions of GRBs by \citet{bromberg2012observational} and \citet{moharana2017observational} could be direct confirmation of theoretical models, they could, with equal plausibility, simply be statistical artifacts. This work is organized as follows: Section \ref{sec:collapsar} reviews and dissects the theoretical arguments for the Collapsar interpretation of the observed plateau of LGRBs duration distribution. Section \ref{sec:statistics} details our alternative statistical interpretation of the plateau and our attempt to reproduce the apparent plateaus seen in observational GRB durations from sample incompleteness. Section \ref{sec:discussion} discusses our results.
	
\section{The Collapsar interpretation of the plateau}
\label{sec:collapsar}

	The argument for the presence of a plateau in the observed duration distribution of LGRBs begins with the assumption that the intrinsic\footnote{Note that \citet{bromberg2012observational} use the keywords `intrinsic' and `observed' interchangeably to represent rest-frame LGRB duration. In this manuscript, `intrinsic' and `observed' exclusively refer to the GRB rest frame and the observer frame on Earth, respectively.} prompt gamma-ray emission duration ($t_\gamma$) depends exclusively on the LGRB central engine activity time ($t_e$) and the jet breakout time ($t_b$) from the stellar envelope,
	\begin{equation}
		\label{eq:durdef}
		t_\gamma =
		\begin{cases}
			0         &\text{if} ~~t_e \leq t_b ~, \\
			t_e - t_b &\text{if} ~~t_e > t_b ~.
		\end{cases}
	\end{equation}
	
	Realistically, as \citet{bromberg2012observational} state, the jet breakout, engine working times, and the gamma-ray emission duration might be correlated with each other and other properties of the LGRB progenitor. Nevertheless, assuming the validity of \eqref{eq:durdef}, one can write the probability density function (PDF) of $t_\gamma$ (i.e., the probability distribution of the intrinsic LGRB duration) in terms of the PDF of the engine working time $t_e$,
	\begin{equation}
		\label{eq:durpdf}
		\pi_\gamma(t_\gamma) \diff t_\gamma =
		\begin{cases}
			0         &\text{if} ~~t_e \leq t_b ~, \\
			\pi_e(t_e) \diff t_e = \pi_e(t_b + t_\gamma) \diff t_\gamma &\text{if} ~~t_e > t_b ~.
		\end{cases}
	\end{equation}
	
	\noindent
	where $\pi$ denotes the PDF. In other words, the duration distribution of LGRBs (for a given $t_b$) is simply the tail of the distribution of the engine work time beyond $t_b$.
	Under further assumption that $\pi_e$ is locally analytic at $t_e = t_b$, the Taylor expansion of the right-hand side of \eqref{eq:durpdf} at $t_e = t_b$ yields,
	\begin{equation}
		\label{eq:taylor}
		\pi_e(t_b + t_\gamma) = \pi_e(t_b) + t_\gamma \frac{\diff \pi_e(t_e)}{\diff t_e}\bigg|_{t_e = t_b} + \orderof \bigg(t_\gamma^2 \bigg) ~.
	\end{equation}
	
	
	Given \eqref{eq:durpdf}, \eqref{eq:taylor} implies a nearly constant PDF for the prompt emission duration of LGRBs as $t_\gamma \rightarrow 0$ if and only if the higher order terms in the Taylor expansion are negligible compared to the first (constant) term. This requires either $t_\gamma \ll t_b$ or all derivatives of $\pi_e(t_e)$ near $t_b$ be nearly zero. Specifically, the second term in the right-hand-side of \eqref{eq:taylor} containing the first derivative of $\pi_e(t_e)$ must satisfy the condition,
	\begin{equation}
		\centering
		\frac{\diff \pi_e(t_e)}{\diff t_e}\bigg|_{t_e = t_b} \ll \frac{\pi_e(t_b)}{t_\gamma} ~. \label{eq:taylorSecTerm}
	\end{equation}
	
	Independent theoretical arguments \citep[e.g.,][]{bromberg2011low} suggest a typical jet breakout time $\hat{t}_b \simeq 50$ [s] which is of the same order as the starting point of the plateau behavior in the duration distribution of LGRBs at $\hat{t}_\gamma \simeq 20 - 30$ [s] \citep{bromberg2012observational}. This similarity ($\hat{t}_\gamma \simeq \hat{t}_b$) clearly violates the condition $t_\gamma \ll t_b$ under which the Taylor expansion is valid. \cite{bromberg2011low, bromberg2012observational} reconcile this by further ``assuming that $\pi_e(t_e)$ is a smooth function and does not vary on short timescales in the vicinity of $\hat{t}_b$", that is, the first and higher-order derivatives in the Taylor expansion must be effectively zero relative to the first constant term in \eqref{eq:taylor}.
	
	Such a strict additional constraint on the derivatives in the Taylor expansion \eqref{eq:taylor} leads to a circular logic where the Collapsar interpretation of the observed nearly-flat plateau in $t_\gamma$ requires the assumption of a nearly-flat plateau in engine working time distribution around $t_b$. Recall that $\pi_\gamma(t_\gamma > 0) = \pi_e(t_e > t_b)$, by definition \eqref{eq:durpdf}. In other words, the Collapsar interpretation of the plateau requires the implicit assumption of the existence of the plateau.
	
	The above circular logic can be resolved in several ways: 1) The theoretical predictions of the typical breakout time $\hat{t} \sim 50$ [s] are imprecise. 2) The distribution of the engine activity time $\pi_e(t_e)$ is indeed nearly flat around $\hat{t}_b$. 3) The observed plateau in the duration distribution of LGRBs does not have a Collapsar interpretation but is due to the statistical nature of the distributions of strictly positive physical quantities combined with sample incompleteness, convolution effects, contamination with SGRBs, and visual effects.
	
	In the following section, we argue the last resolution offers a plausible explanation for the apparent plateaus in the duration distributions of both LGRBs and SGRBs without invoking any physical theories.

\section{The statistical interpretation of the plateau}
\label{sec:statistics}

	\begin{figure*}
		\centering
		\makebox[\textwidth]
		{
			\begin{tabular}{cc}
				\subfloat[]{\includegraphics[width=0.47\textwidth]{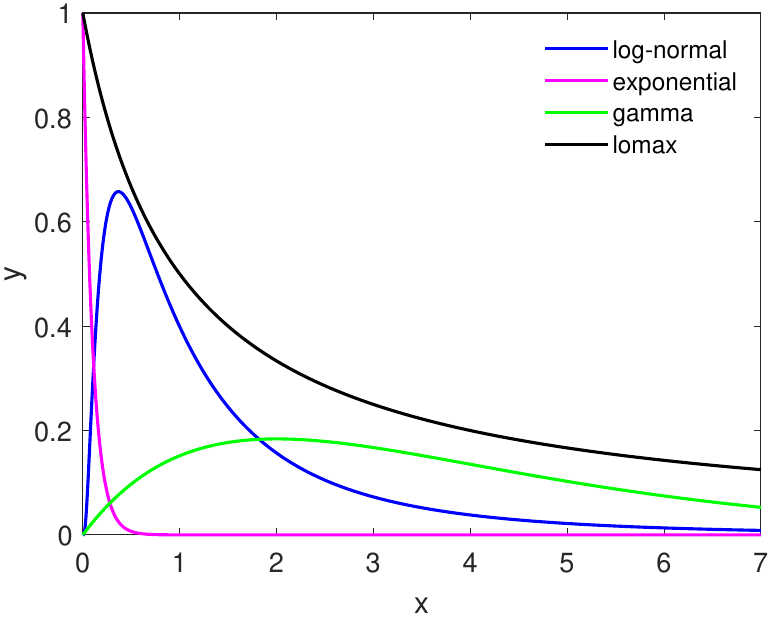} \label{fig:taylorZoomOut}} &
				\subfloat[]{\includegraphics[width=0.47\textwidth]{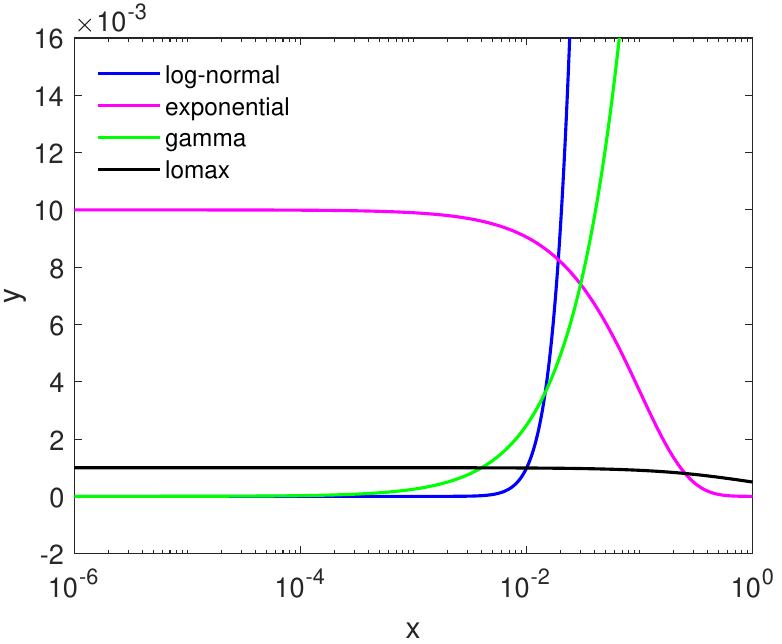} \label{fig:taylorZoomIn}}
			\end{tabular}
		}
		\caption{\label{fig:taylor}
			An illustration of the plateau in the distribution of strictly positive random variables as $x \rightarrow 0$. The existence of these plateaus is mathematically guaranteed for all distributions with positive support and finite-valued Probability Density Function (PDF).
			{\bf (a)} The PDFs of three popular positive-valued statistical distributions.
			{\bf (b)} A zoom-in on the same PDFs as in plot (a), but at very small values near the origin, on a logarithmic x-axis, illustrating the plateau-like behavior of the distributions near $x = 0$. The appearance of the plateau near the origin ($x = 0$) is guaranteed by the logarithmic transformation of the x-axis. The logarithmic transformation effectively spreads a finite amount of variations in the PDF on the y-axis over a semi-infinite range on the (logarithmic) x-axis.
		}
	\end{figure*}
	
	Statistical distributions with strictly positive support, for example, $\pi(t_\gamma), t_\gamma \in (0,+\infty)$, frequently and naturally exhibit plateaus in their short tails. Such plateaus can appear under different independent circumstances discussed in the following subsections.
	
	\subsection{All finite-valued statistical distributions with positive support have plateaus}
	\label{sec:statistics:taylor}
	
	Recall that the Taylor expansion of the PDF of an analytic statistical distribution at any point within its support guarantees the existence of a plateau near the point of expansion. Such a plateau, however, is mathematically infinitesimal, defined only asymptotically as one approaches the point of expansion. Therefore, these mathematically infinitesimal plateaus are invisible to the human eye almost anywhere within the support of the PDF, except at the origin $t_\gamma = 0$ under an appropriate transformation. By definition, the PDF has a positive support $t_\gamma\in(0,+\infty)$. Therefore, taking the logarithm of the x-axis ($t_\gamma$) pushes the lower limit of the support of the PDF at $t_\gamma = 0$ to negative infinity:  $\log(t_\gamma\rightarrow0) \rightarrow -\infty$. This logarithmic transformation of the x-axis effectively infinitely magnifies the Taylor expansion of $\pi(t_\gamma)$ around $t_\gamma = 0$. This infinite magnification can be intuitively understood by noting that the PDF $\pi(t_\gamma)$ of the distribution must have a finite value at the origin $t_\gamma = 0$ with a well-defined finite right-hand derivative. When the x-axis is logarithmically transformed, the finite amount of changes in $\pi(t_\gamma)$ span progressively over larger and larger logarithmic ranges on the x-axis, effectively making the PDF look like a plateau as $\log(t_\gamma)\rightarrow-\infty$.
	
	This is precisely the mechanism by which the Collapsar interpretation discussed in Sect \ref{sec:collapsar} attempts to explain the observed plateau in the duration distribution of LGRBs. However, the Collapsar theory of LGRBs invocation appears unnecessary since all finite-valued statistical distributions with strictly positive support exhibit a plateau toward zero, and there is an uncountably infinite number of such statistical distributions. Given that the gamma-ray duration $t_\gamma$ of both SGRBs and LGRBs is a strictly positive-valued observable random variable, the appearance of a plateau in their $t_\gamma$ distributions on a $\log(t_\gamma)$-axis is mathematically guaranteed as $t_\gamma\rightarrow0$. This is true without recourse to any physical theories of GRBs. Figure \ref{fig:taylor} illustrates this mathematical asymptotic plateau behavior for some well-known continuous distributions with positive real support.
	
	\subsection{Sample-incompleteness creates plateaus in observational data}
	\label{sec:statistics:sample}
	
	Despite the mathematical guarantee of a plateau in strictly positive finite-valued distributions, the practical visibility of such plateaus in observational data is limited to distributions whose probability mass is heavily concentrated around zero. For an illustration, the near-zero, hard-to-sample, mathematical plateaus of Gamma and Lognormal distribution from this category are depicted in Figure \ref{fig:taylorZoomIn}.
	

	Sample-incompleteness, however, can still generate a second kind of plateau in strictly positive statistical distributions, entirely different from the plateaus of the first kind of mathematical origin discussed in Sect \ref{sec:statistics:taylor}. Such observational plateaus result from limited sampling of positive statistical distributions that are highly positively skewed. Recall that plateaus naturally also occur in the neighborhood of the mode of a distribution where the derivative of the PDF is mathematically zero. 
	
	This flatness of the PDF around the mode can readily become observationally visible if the distribution sharply rises from some finite value at $t_\gamma = 0$, typically $\pi(t_\gamma = 0) = 0$, to the PDF mode at $t_\gamma = \hat{t}$, and gradually declines to zero as $t_\gamma\rightarrow+\infty$. This sharp rise and gradual decay is the typical behavior of many positive-valued statistical distributions and the key requirement for generating finite-sample extended plateaus in their PDFs.
	
	Therefore, if the shape of the distribution (in log-log space) is such that the probability of sampling at $t_\gamma < \hat{t}$ is negligible, a plateau appears in the histogram of observational data at $t_\gamma \sim \hat{t}$. The plateau appearance is substantially strengthened and extended if the PDF decays slowly (and concavely) toward $+\infty$ and the observational sample is binned and visualized on a logarithmic x-axis as done in \citet{bromberg2012observational}.
	
\begin{figure*}
		\centering
		\makebox[\textwidth]
		{
			\begin{tabular}{llll}
				\subfloat[]{\includegraphics[width=0.23\textwidth]{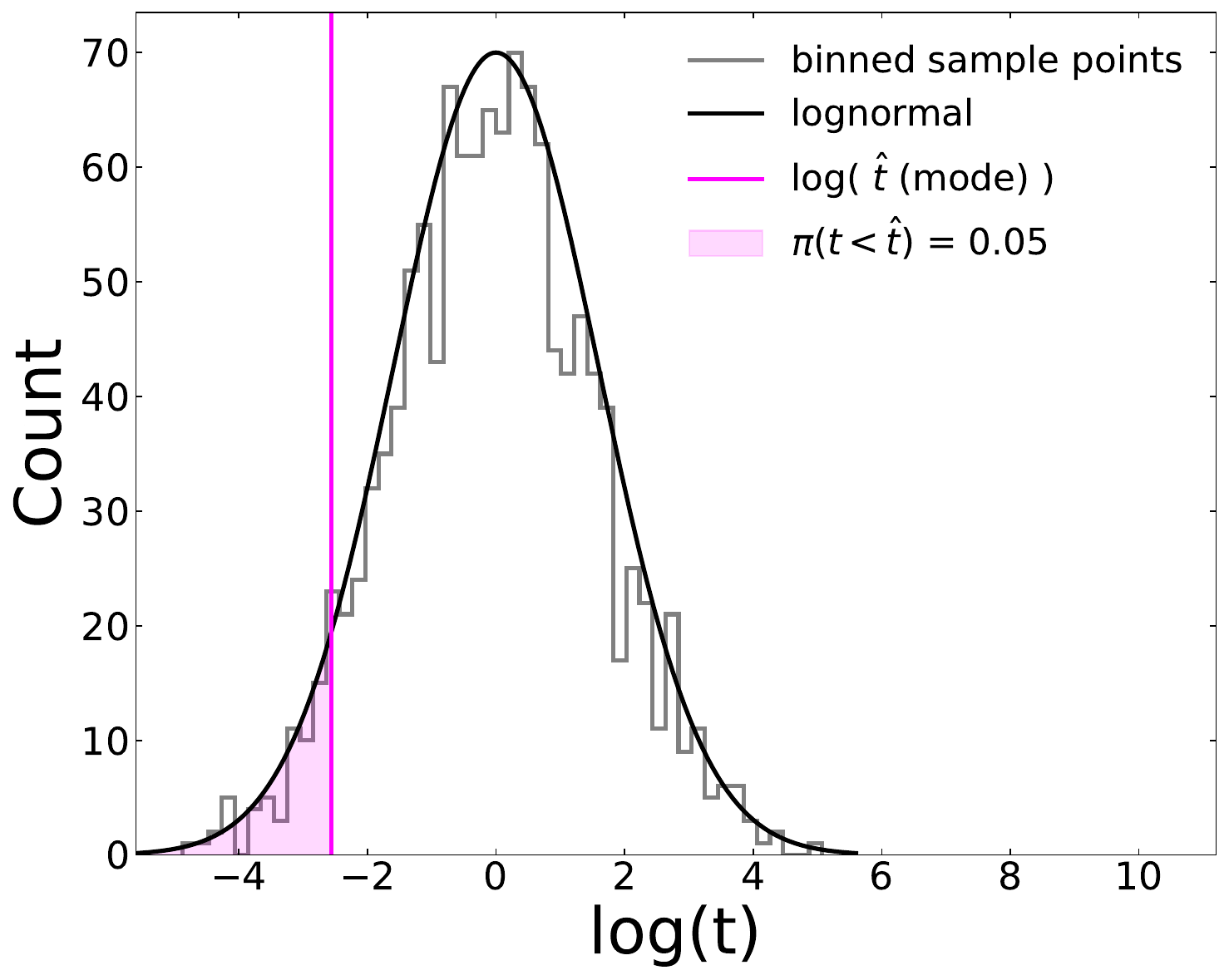} \label{fig:lognormalA}}
				&
				\subfloat[]{\includegraphics[width=0.23\textwidth]{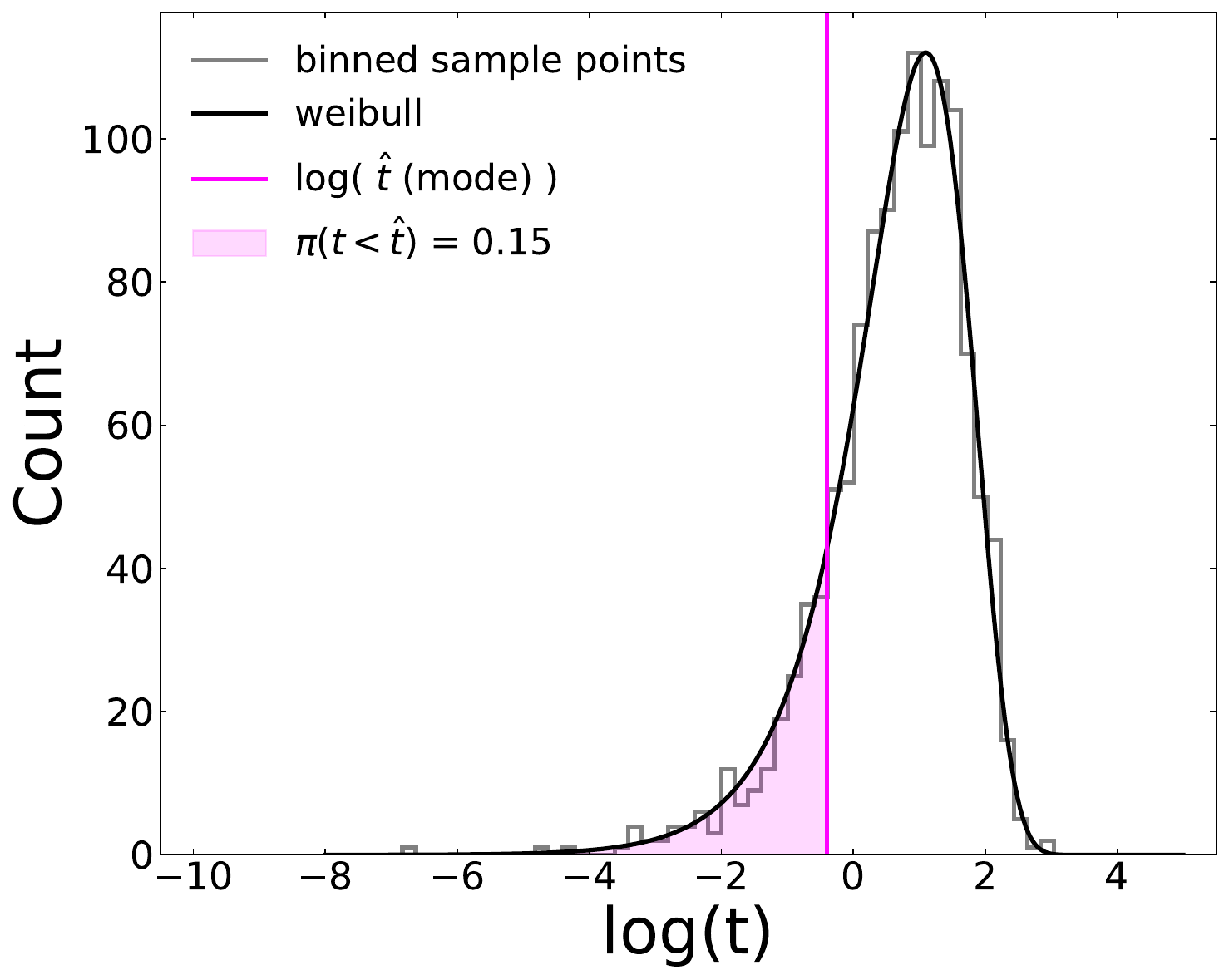} \label{fig:weibullA}}
				&
				\subfloat[]{\includegraphics[width=0.23\textwidth]{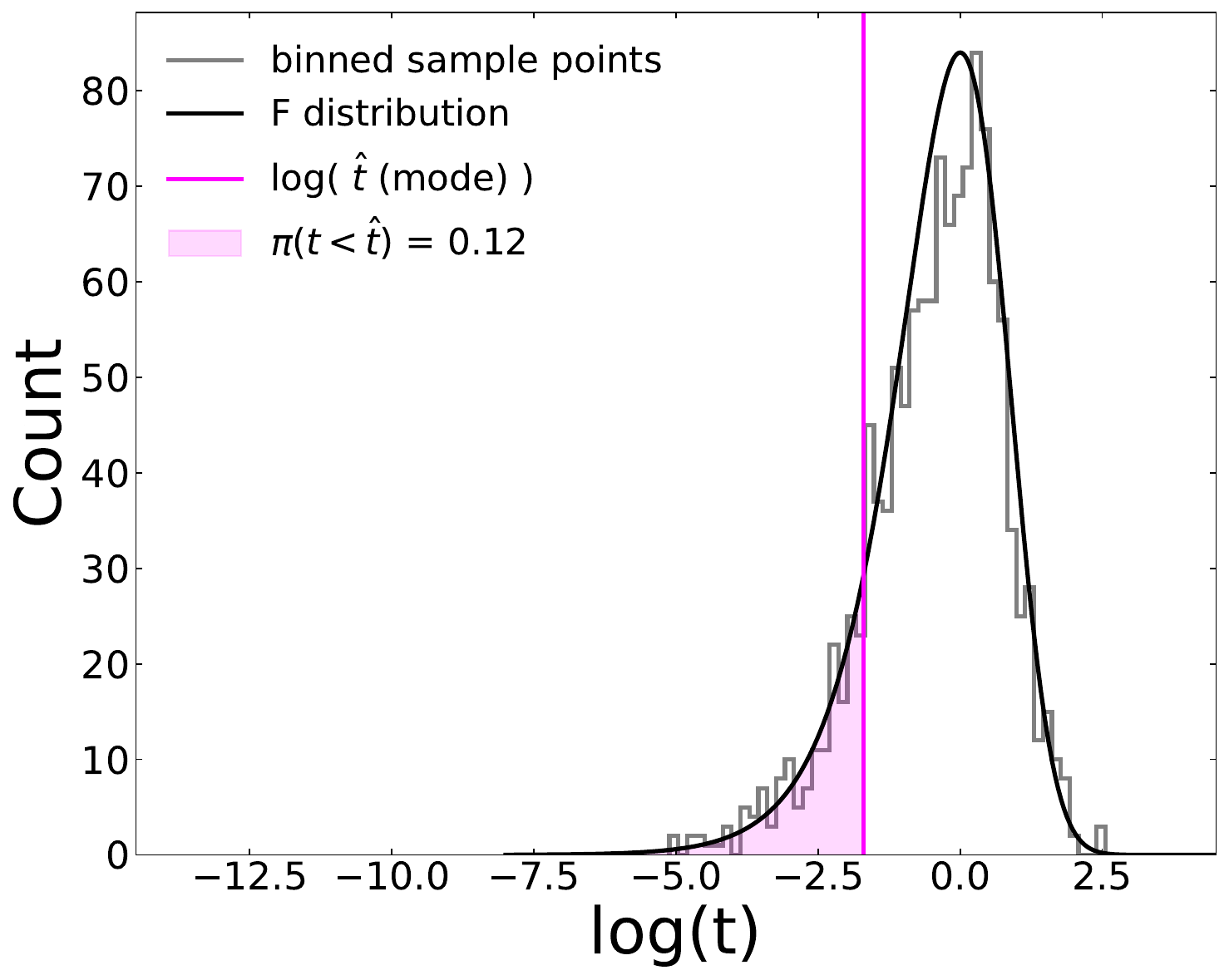} \label{fig:fA}}
				&
				\subfloat[]{\includegraphics[width=0.23\textwidth]{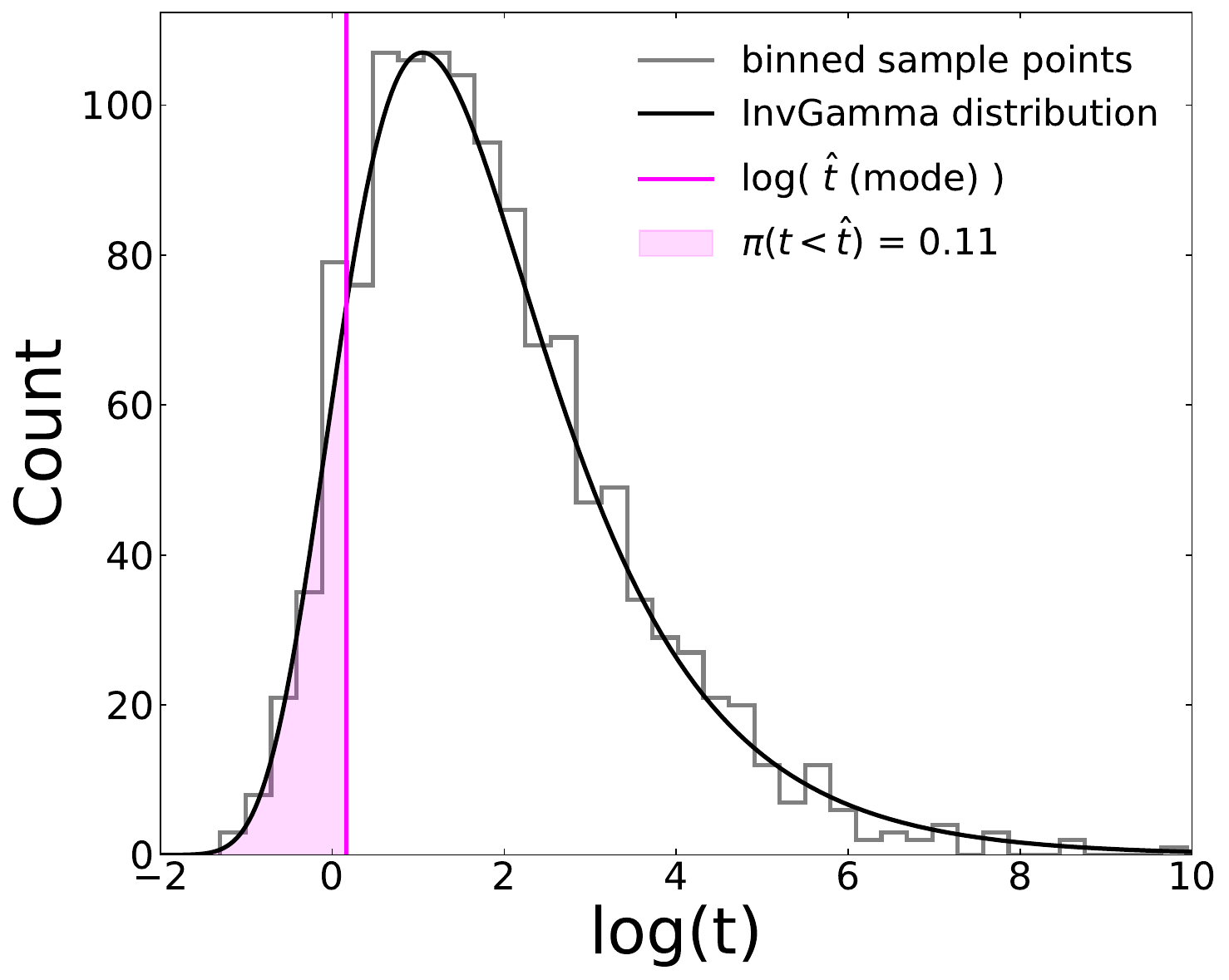} \label{fig:invgammaA}}
				\\
				\subfloat[]{\includegraphics[width=0.23\textwidth]{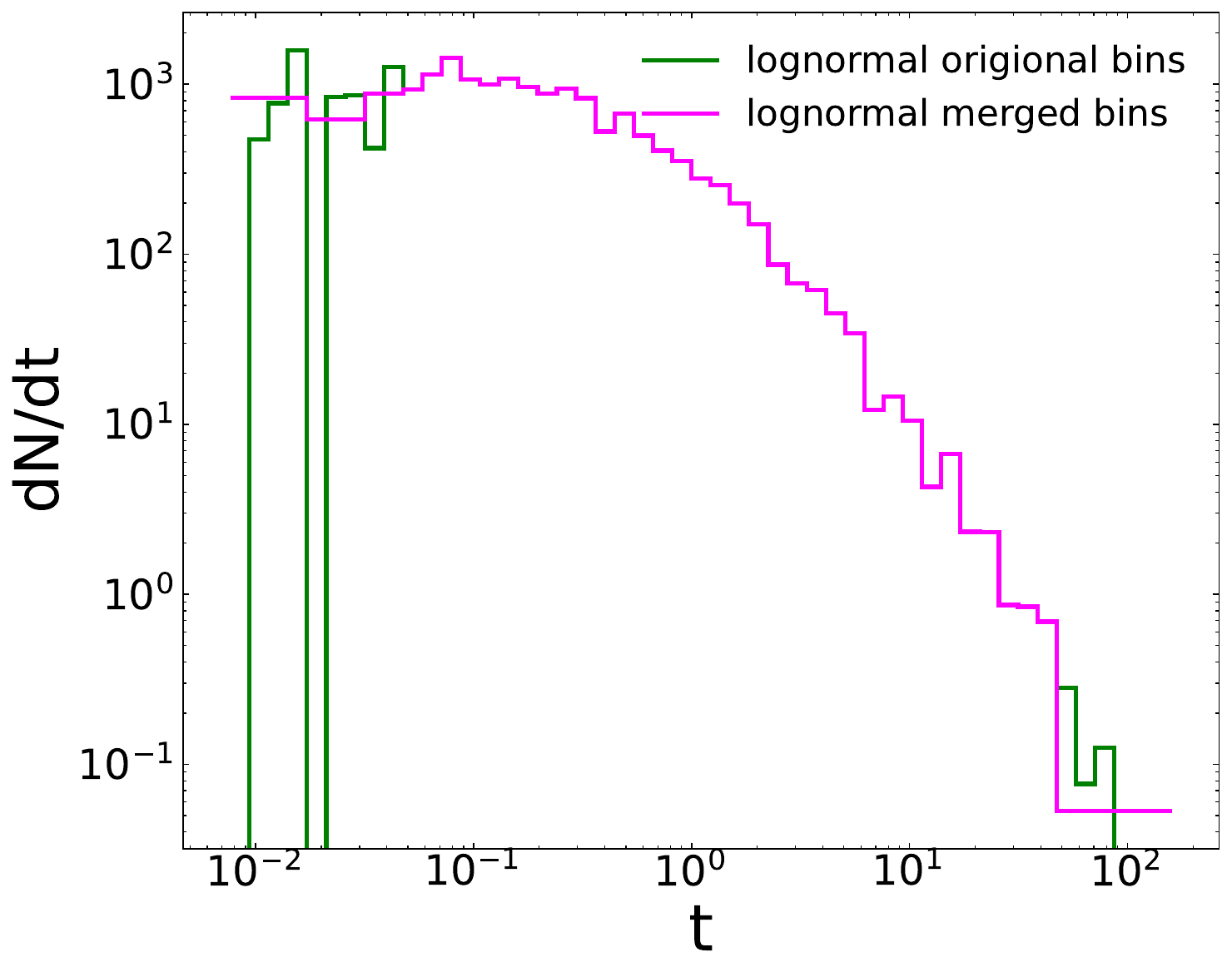} \label{fig:lognormalB}}
				&
				\subfloat[]{\includegraphics[width=0.23\textwidth]{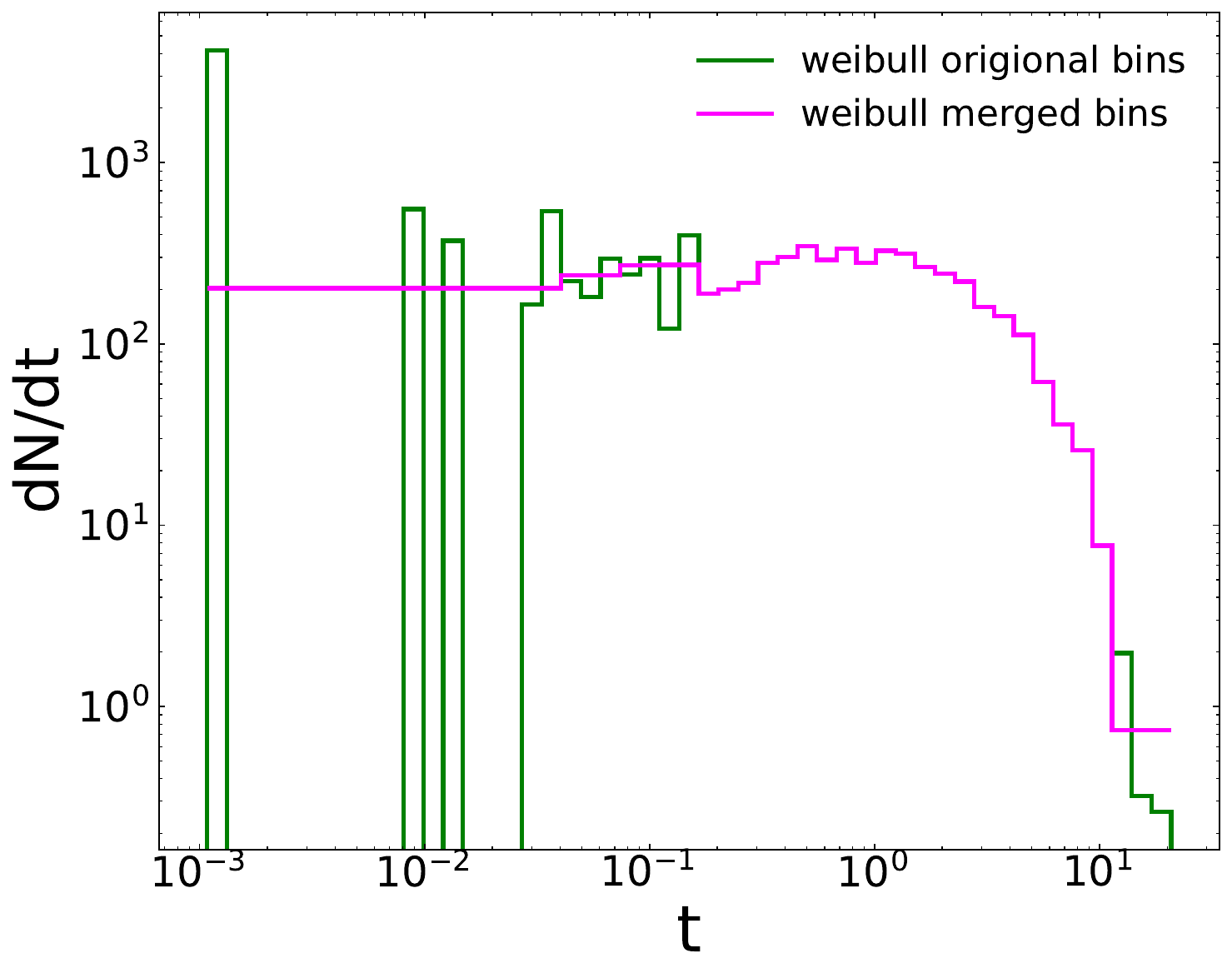} \label{fig:weibullB}}
				&
				\subfloat[]{\includegraphics[width=0.23\textwidth]{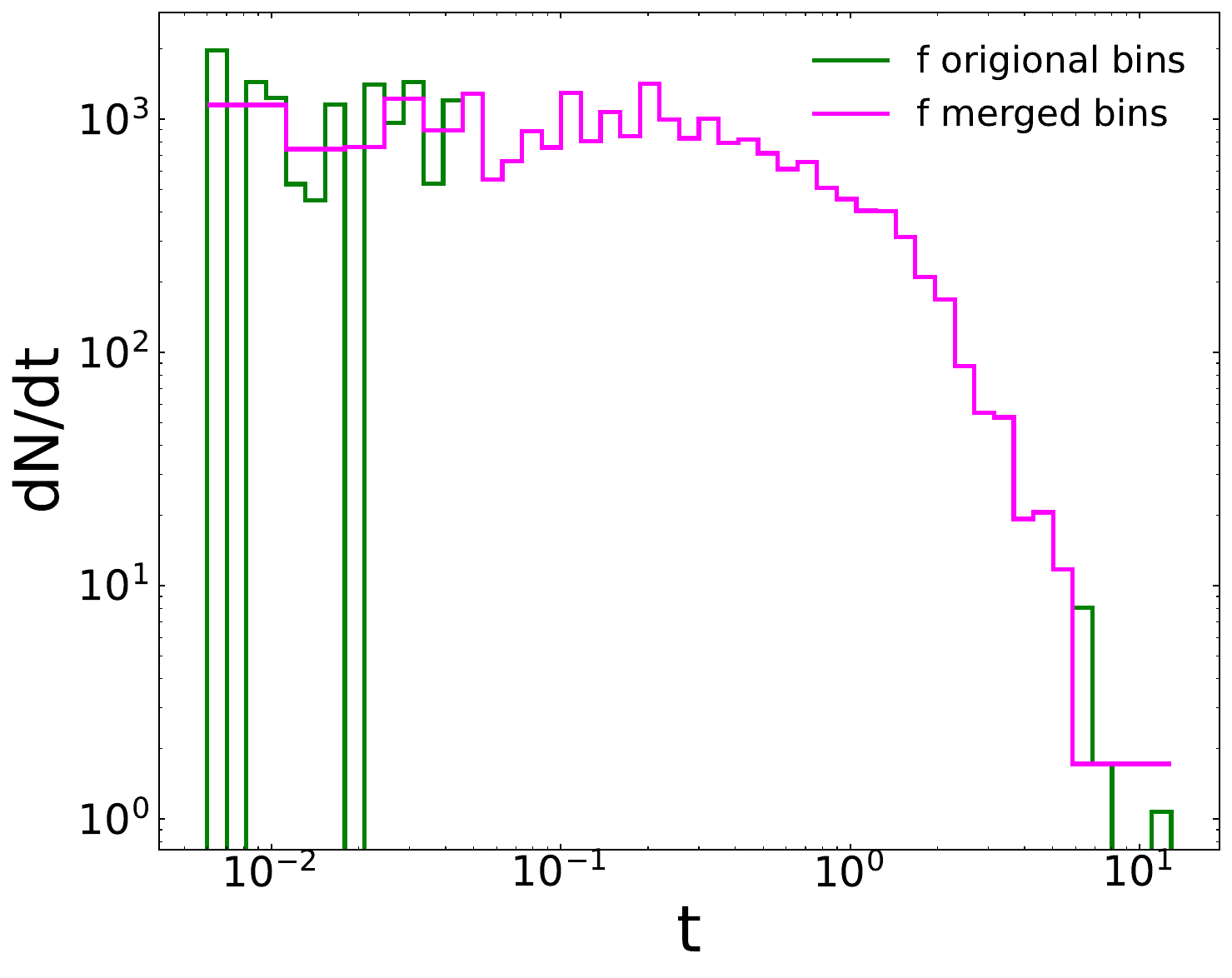} \label{fig:fB}}
				&
				\subfloat[]{\includegraphics[width=0.23\textwidth]{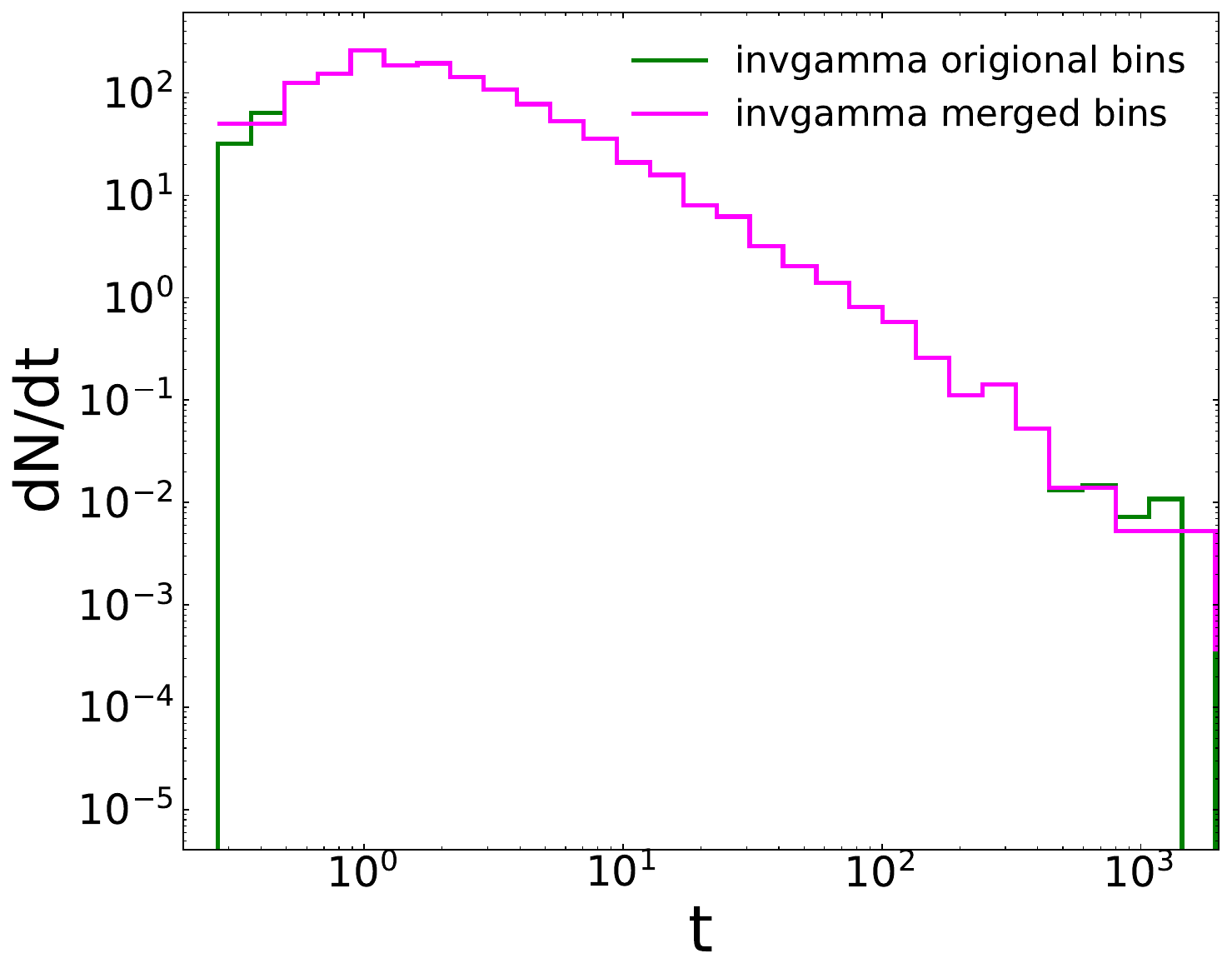} \label{fig:invgammaB}}
			\end{tabular}
		}
		\caption{
			An illustration of the effects of sample incompleteness on plateau formation in statistical distributions. A primary factor in the appearance of finite-sample plateaus near the modes of PDFs is the area to the left of the PDF mode. The smaller this area is, the fewer opportunities there are to construct the short tail of the PDFs in finite sampling, thus leading to the appearance of plateaus in the PDFs near their modes. The top plots depict the histograms of the log-transformed random variables drawn from select statistical distributions with positive support. The bottom plots depict the corresponding histograms of the original random variables but on logarithmic axes.
			\label{fig:saminc}
		}
	\end{figure*}

	To illustrate this artificial finite-sample plateau creation in action, consider the Lognormal distribution, widely used in Astronomy for modeling luminosity functions and other observational data \citep[e.g.,][]{balazs2003difference, butler2010cosmic, berkhuijsen2012density, 2013arXiv1308.1097S, shahmoradi2013multivariate, shahmoradi2015short, 2017arXiv171110599S, medvedev2017statistics, zaninetti2018lognormal, osborne2020multilevelapj, yan2021characteristic}. The Lognormal distribution is always positively skewed for any parameter values. This results in a typically negligible area to the left of the PDF mode, as illustrated in Figures \ref{fig:lognormalA} and \ref{fig:lognormalB}. Hence, the short tail of the Lognormal distribution to the left of its mode is rarely fully observationally constructed, leading to the appearance of a plateau in the PDF on the log scale near the distribution mode. This behavior is not exclusive to Lognormal and is generically seen in many statistical distributions with positive support, some of which are also shown in the rest of the plots of Figure \ref{fig:saminc}.
	
	\subsection{Convolution creates plateaus in observational data}
	\label{sec:statistics:conv}
	
	In addition to the area to the left of the mode of PDFs with positive support, the smoothness (differentiability) and concavity of the PDF near its mode play a dominant role in the appearance of finite-sample plateaus.
	Indeed, a separate class of statistical distributions with strictly positive support has modes that are neither smooth nor occur at the origin. These distributions, exemplified by the Pareto and Log-Laplace, hardly exhibit plateau behavior under finite sampling. Nevertheless, this section shows that even the most difficult, non-analytic, non-smooth PDFs can exhibit plateaus under convolution.
	
	Recall the convolution of two functions,
	\begin{equation}
		\label{eq:conv}
		(f * g)(t) = \int_{-\infty}^{\infty} f(t)g(t-\tau) \,d\tau ~,
	\end{equation}

	\noindent
	acts as a smoothing operation similar to that of weighted averaging. The resulting function from the convolution will be at least as smooth as the individual functions ($f(\cdot)$ or $g(\cdot)$ in \eqref{eq:conv}). In fact, the convolution operation is identical to a weighted dynamic averaging of $f(\cdot)$ when the weight, $g(\cdot)$, is a probability density function. In the case of the duration distribution of LGRBs, the convolution performed to obtain the observer-frame duration distribution is the following,
	\begin{equation}
		\label{eq:convz}
		\log(t_{\gamma, obs}) = \log(z + 1) + \log(t_{\gamma, int}) ~,
	\end{equation}
	
	\noindent
	where ``obs" and ``int" stand for the observed and intrinsic durations, respectively, and z represents GRB redshift.
	
	\begin{figure*}
		\centering
		\makebox[\textwidth]
		{
			\begin{tabular}{llll}
				\subfloat[Pareto for intrinsic $t_{\gamma,\mathrm{LGRBs}}$]{\includegraphics[width=0.2325\textwidth]{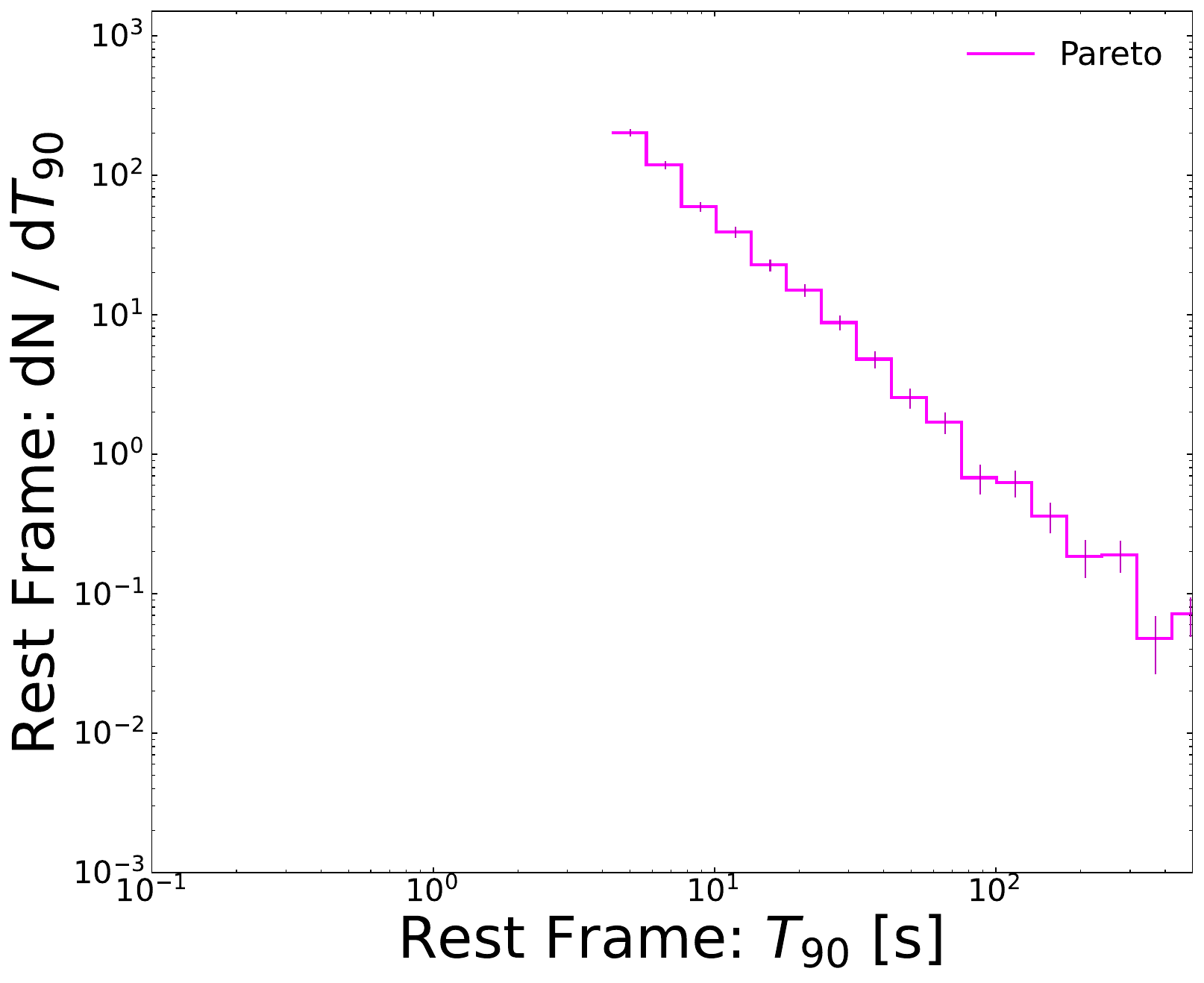} \label{fig:paretoalone}}
				&
				\subfloat[Observed redshift-convolved $t_{\gamma,\mathrm{LGRBs}}$]{\includegraphics[width=0.2325\textwidth]{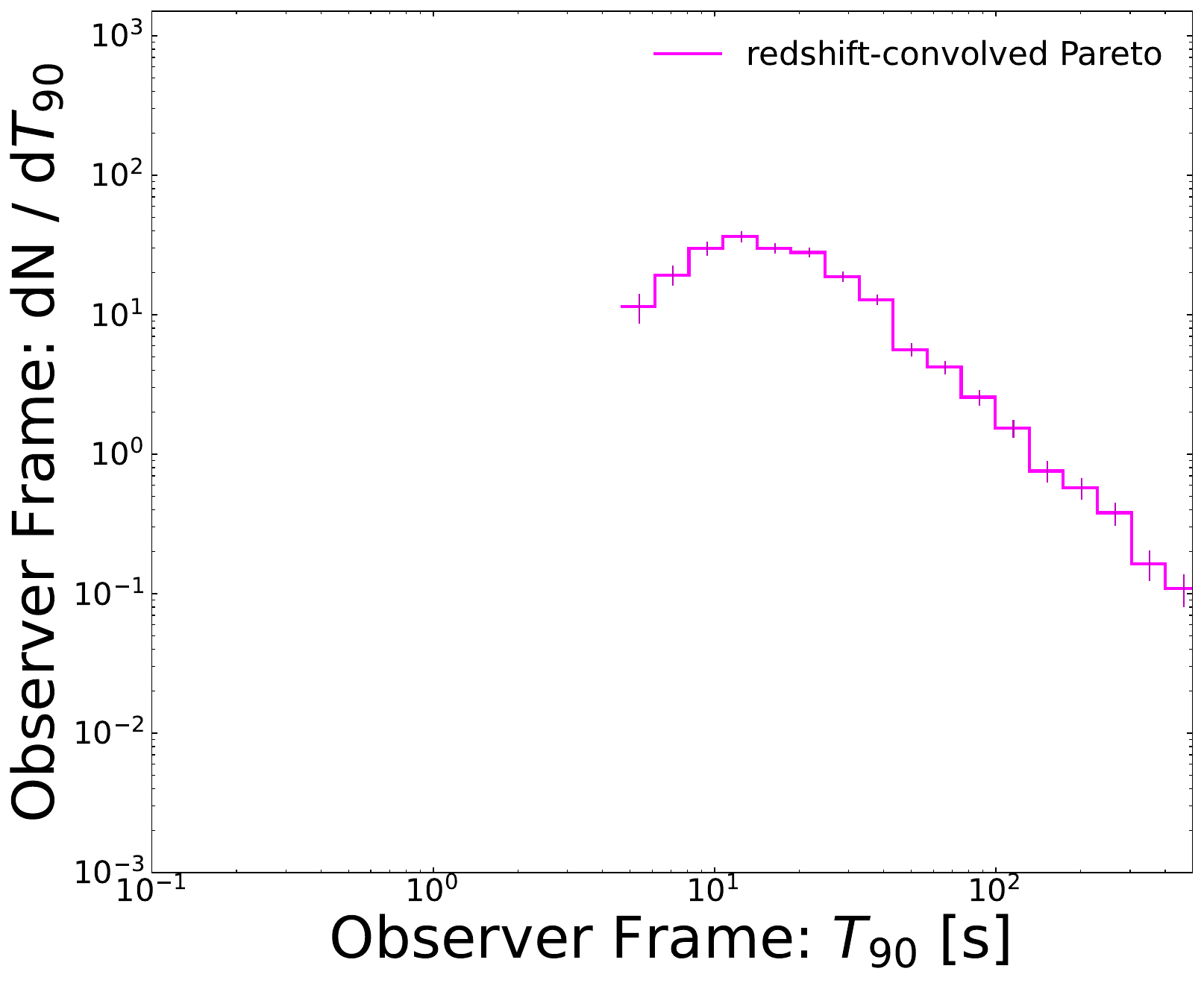} \label{fig:paretoconv}}
				&
				\subfloat[Mixture of $t_{\gamma,\mathrm{LGRBs}}$ and $t_{\gamma,\mathrm{SGRBs}}$]{\includegraphics[width=0.2325\textwidth]{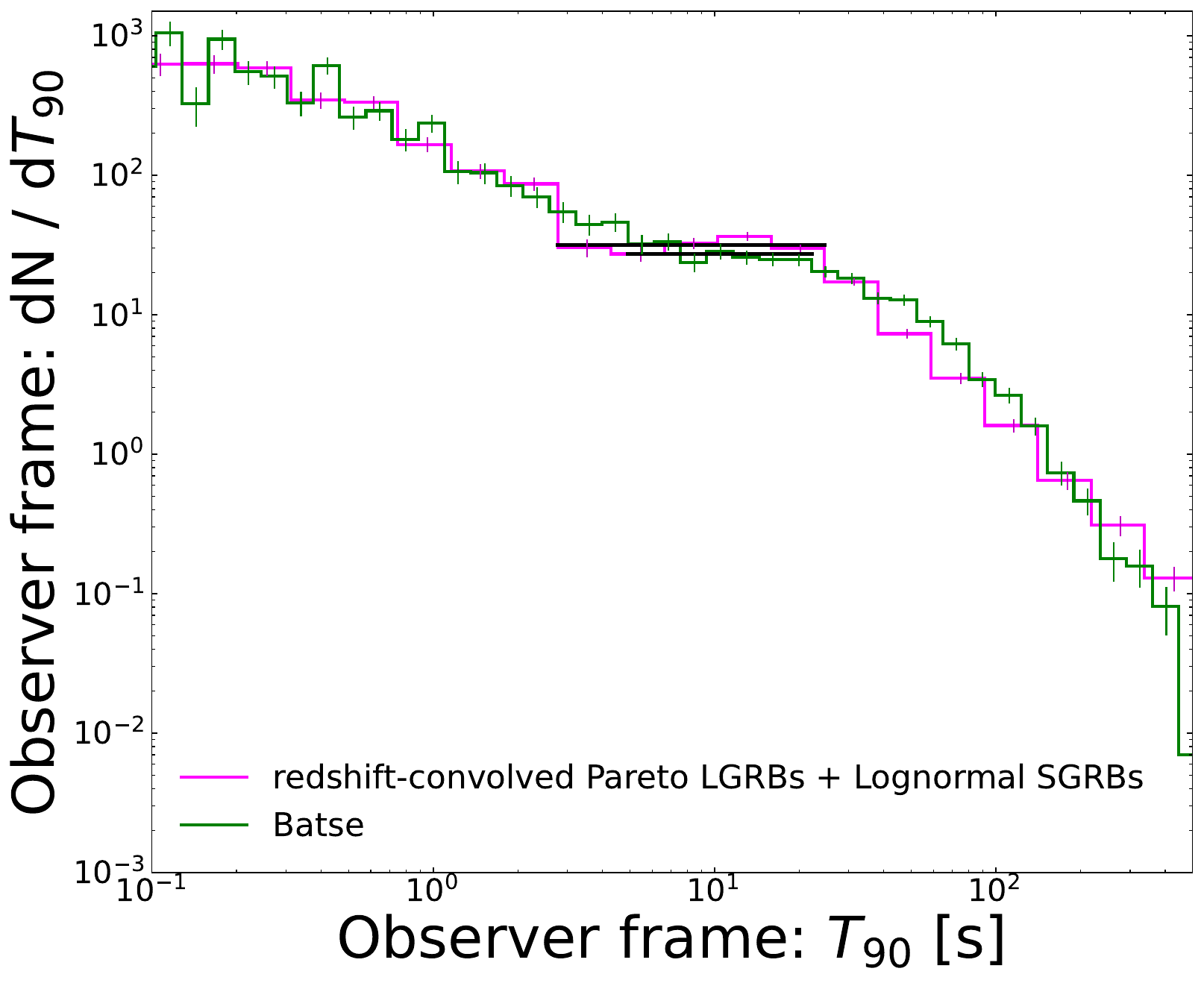} \label{fig:paretoconvmixed}}
				&
				\hspace*{-5mm}
				\subfloat[distribution of plateau lengths]{\includegraphics[width=0.26\textwidth]{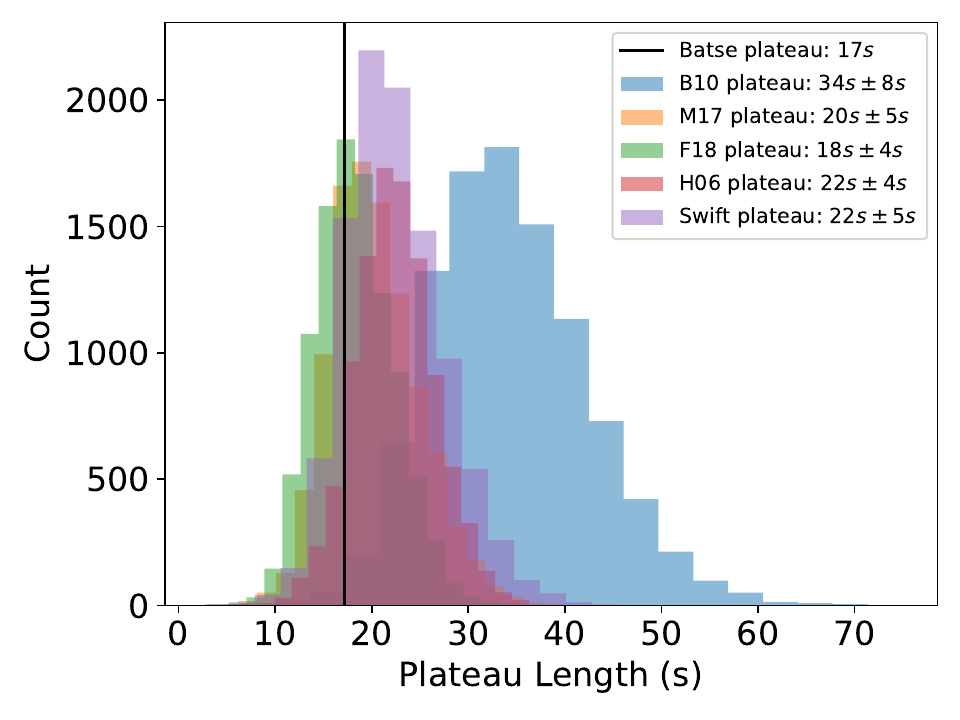} \label{fig:histParetoPlateauLength}}
			\end{tabular}
		}
		\caption{
			An illustration of the smoothing effects of convolution on the Pareto distribution, leading to the appearance of a plateau in $\frac{dN}{dT}$ plot. \textbf{Plot (a)} shows sampled points from a Pareto distribution after performing the bin merging as described in \citet{bromberg2012observational}. \textbf{Plot (b)} shows the same data as in plot (a) but convolved with the redshift distribution of LGRBs (derived from Swift catalog). \textbf{Plot (c)} further combines the convolved observed duration distribution of LGRBs with a Lognormal fit to the redshift distribution of SGRBs. For comparison, the green line represents the duration distribution of the BATSE SGRBs and LGRBs. The final observed plateau resulting from the Pareto distribution is $50\%$ longer than the observed plateau in the duration distribution of BATSE LGRBs. \textbf{Plot (d)} illustrates the distribution of the average plateau lengths assuming an intrinsic Pareto duration distribution convolved with various GRB popular redshift distribution scenarios from the literature as listed in the plot legend. The observed plateau length in the BATSE duration data is illustrated by the vertical black line.
			\label{fig:pareto}
		}
	\end{figure*}
	
	We demonstrate the effectiveness of convolution in creating plateaus by considering a Pareto PDF for the duration distribution of LGRBs. The Pareto distribution does not exhibit any plateau-like behavior within its support under any circumstances (e.g., see Figure \ref{fig:paretoalone}). Yet, when convolved with the redshift distribution of LGRBs detected by the Neil Gehrels Swift Observatory\footnote{\href{https://swift.gsfc.nasa.gov/results/batgrbcat/summary_cflux/summary_general_info/GRBlist_redshift_BAT.txt}{All redshifts are taken from the Swift GRB catalog.}}, the sharp non-analytical mode of the Pareto PDF transforms into a continuous smooth mode as seen in Figure \ref{fig:paretoconv}. Combining the observed convolved duration distribution of LGRBs with the duration distribution of SGRBs further eliminates the appearance of any discontinuity or sharp decline in the short tail of the distribution, leading to a plateau in the mixture distribution that is even longer than the observed plateaus in the duration distributions of The Burst And Transient Source Experiment (BATSE) and Fermi catalog GRBs (Figure \ref{fig:paretoconvmixed}). 
	
	To verify the generality of convolution effects on plateau creation, we also consider four additional popular GRB formation rate scenarios, each in 10,000 independent simulations: \citet{fermi2018gamma} (F18), \citet{madau2017radiation} (M17), \citet{butler2010cosmic} (B10), \citet{hopkins2006normalization} (H06). Data modeling and fittings are performed by assuming one of these redshift models as the redshift distribution of LGRBs and assuming a parametric form for the intrinsic duration distribution of LGRBs (e.g., Pareto), using a Bayesian fitting approach as described in \citet{shahmoradi2013multivariate, shahmoradi2015short, 2017arXiv171110599S, osborne2020multilevelapj} assuming non-informative objective priors for all unknown parameters. Once the model fitting is carried out, random redshift samples of the same size as the observational duration distribution data are generated by sampling the corresponding redshift distribution models with an adaptive Markov Chain Monte Carlo algorithm \citep{Shahmoradi2020a, shahmoradi2020paradram, 2020arXiv201000724S, Kumbhare2020, shahmoradi2020paramonteII}. The random samples are subsequently convolved with the random samples from the corresponding duration distribution models with the best-fit parameters to generate the observed duration distributions. This procedure is repeated 10,000 times for each modeling and redshift distribution scenario. Finally, the lengths of the resulting apparent plateaus in the duration distributions for each of the 10,000 simulations per modeling scenario are measured using the same approach as in \citet{bromberg2012observational}. 
	
	The histograms of the resulting plateau lengths for each of the five redshift distribution scenarios are illustrated in plot (d) of Figure \ref{fig:histParetoPlateauLength}. Notably, the resulting average plateau lengths from the convolution of a Pareto duration distribution with all GRB formation rate scenarios is longer than the plateau length in BATSE observational data.
	
	\subsection{Sample contamination creates plateaus in observational data}
	\label{sec:statistics:contamination}
	
\begin{figure*}
		\centering
		\makebox[\textwidth]
		{
			\begin{tabular}{ccc}
				\subfloat[]{\includegraphics[width=0.31\textwidth]{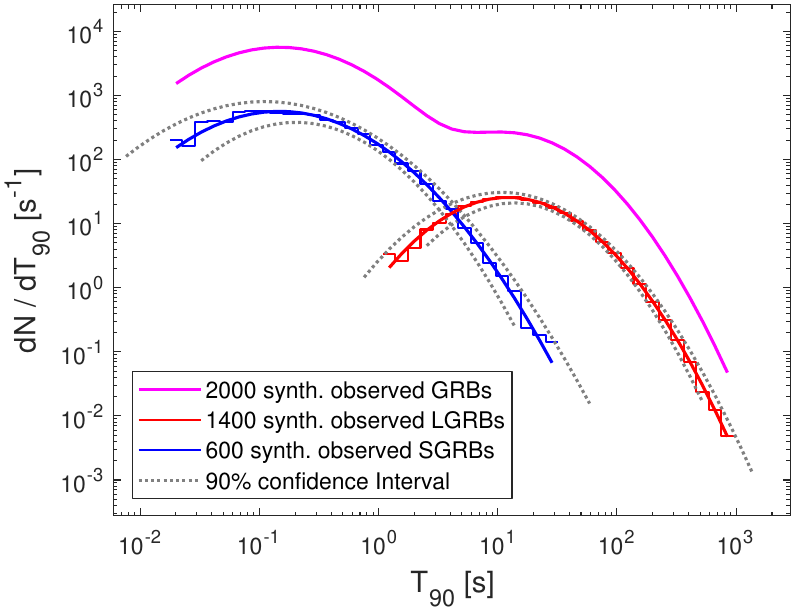} \label{fig:allDNDTObserved}} &
				\subfloat[]{\includegraphics[width=0.31\textwidth]{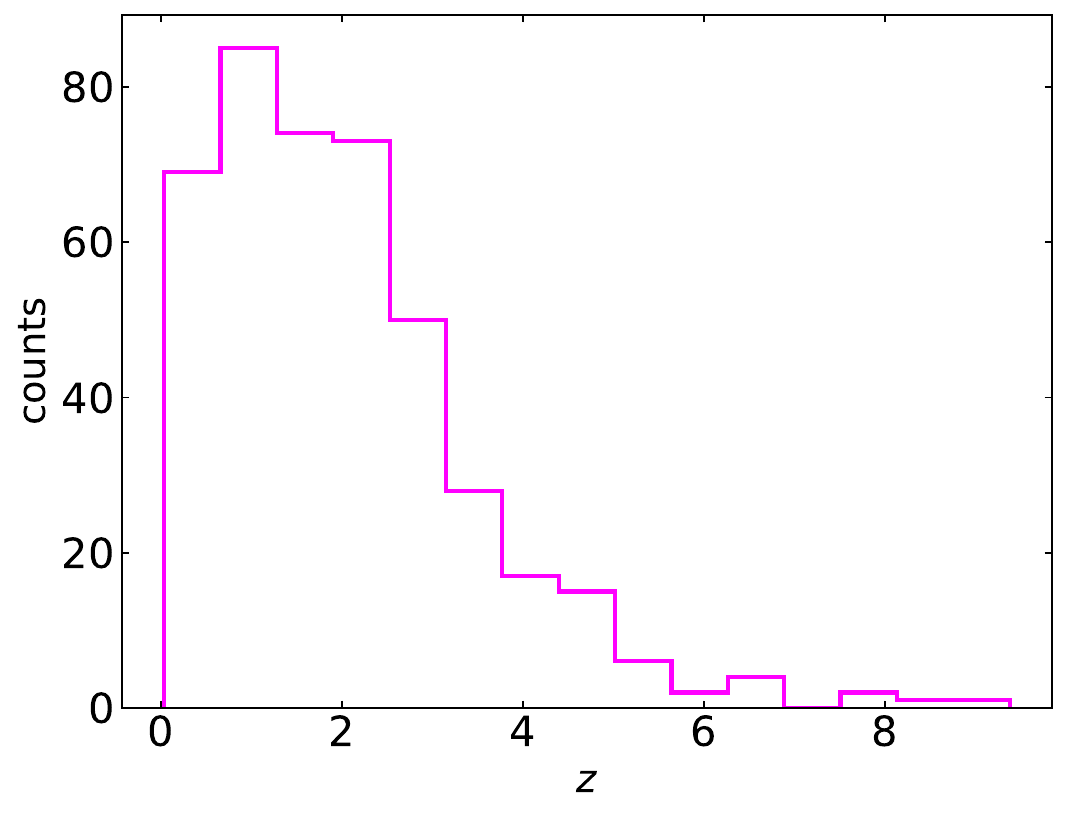} \label{fig:z}} &
				\subfloat[]{\includegraphics[width=0.31\textwidth]{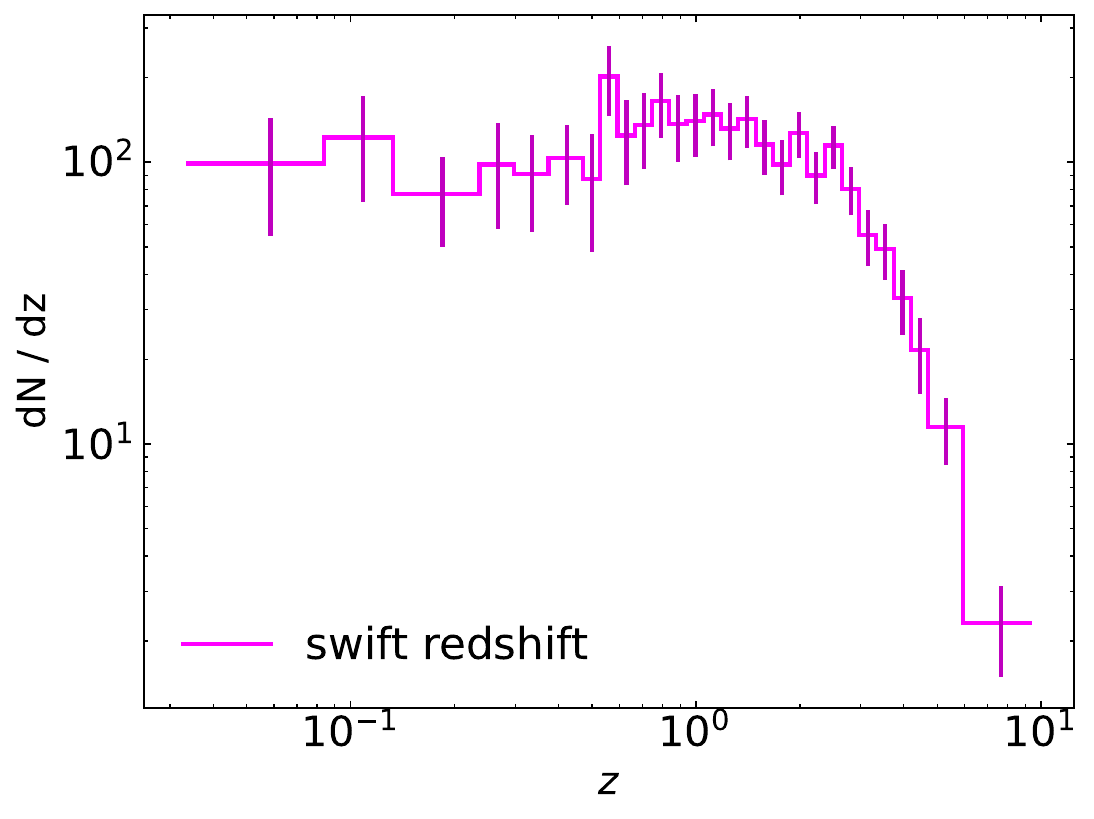} \label{fig:zDnDz}}
			\end{tabular}
		}
		\caption{
			BATSE-detectable samples of 600 SGRBs and 1400 LGRBs are randomly generated from our respective Monte Carlo universes. In plot \textbf{(a)}, SGRBs are represented by the blue bins and LGRBs by the red bins, where each bin is the 50th percentile value of 10,000 synthetic detections. The solid blue and red lines are lognormal fits to the bins in the pre-transformed space. These two fits are summed to produce the solid magenta line multiplied by a factor of 10 to offset it for clarity. The dotted gray lines represent the 90\% confidence interval for each binned distribution. Plot \textbf{(b)} shows the distribution of observed Swift redshifts in linear space. Lastly, plot \textbf{(c)} shows the redshift distribution transformed in the same manner as the duration distribution throughout this manuscript. After performing the transformation described in \citet{bromberg2012observational}, the resulting plot displays a plateau in the log-log space, although there is no apparent physical origin.
			\label{fig:observedGRB}
		}
	\end{figure*}

	Contamination of the short tail of the duration distribution of LGRBs with SGRBs leads to further extension of the apparent plateaus because the population of SGRBs compensates for any drop in the count of LGRBs toward low durations. Again, consider a Lognormal fit for the intrinsic duration distribution of LGRBs. \citet{shahmoradi2013multivariate,2013arXiv1308.1097S,shahmoradi2015short,osborne2020multilevelapj} argue and provide evidence for the goodness of fit of LGRBs and SGRBs prompt duration distributions with Lognormal PDF. The Lognormal distribution does not exhibit an inherent plateau near its short tail. Sample incompleteness, however, creates the appearance of a plateau in the Lognormal fit to the observational LGRB data. This apparent plateau is further extended by the convolution of the intrinsic duration distribution with redshift to obtain the observe-frame duration distribution as illustrated in Figure \ref{fig:allDNDTObserved}. Finally, contamination with SGRBs data due to the overlap of the two distributions completely eradicates any signs of decline in the short tail of LGRBs duration distribution, yielding a perfect plateau appearance in the final mixture distribution as seen in the magenta solid curve in Figure \ref{fig:allDNDTObserved}.
	
\section{Discussion}
\label{sec:discussion}

	In this manuscript, we presented statistical arguments that strongly favor a non-physical (i.e., non-collapsar) origin for the observed plateau in the duration distribution of LGRBs. The Taylor expansion \eqref{eq:taylor} of the density function of the engine activity time ($t_e$) near the jet breakout time ($t_b$) requires the plateau to appear at durations orders of magnitude smaller than the inferred $t_b$ in \citet{bromberg2012observational}. \cite{bromberg2011low, bromberg2012observational} reconcile this inconsistency by an additional assumption that the engine activity time ``is a smooth function and does not vary on short timescales in the vicinity of the breakout time". This, however, creates a circular logic in the argument for the Collapsar origin of the observed plateau, where \cite{bromberg2012observational} impose a flatness condition on the distribution of the engine activity time near the breakout time to obtain a flatness in the prompt duration of LGRBs, which is by definition \eqref{eq:durpdf} the same as the engine activity time at $t_\gamma > t_b$. Alternatively, we can interpret the plateau in the duration distribution of LGRBs as a constraint on the shape of the engine activity time, under the assumption that the prompt gamma-emission duration precisely reflects the engine activity time at $t_\gamma > t_b$, as proposed in \cite{bromberg2011low, bromberg2012observational}. In either case, the plateau in the duration distribution of LGRBs does not appear to serve as an observational imprint of the Collapsar model.
	
	We further question any physical origins of the observed plateau by showing that plateaus are ubiquitous in the short tails of statistical distributions with strictly positive support (e.g., Figure \ref{fig:saminc}) and frequently result from a combination of sample incompleteness with the highly positively-skewed nature of such distributions (on natural axes). Even where the intrinsic duration distribution of LGRBs does not exhibit a plateau behavior in its short tail, we show that its convolution with a redshift distribution can create observed duration distributions that exhibit plateau-like behavior. The presence and extent of the plateaus are further significantly enhanced if SGRBs mix with and {\ contaminate} the observed duration distribution of LGRBs, as is the case with all major GRB catalogs. An example of a perfect plateau resulting from such contamination is depicted in Figure \ref{fig:allDNDTObserved}.
	
\begin{figure}
		\centering
		\includegraphics[width=0.47\textwidth]{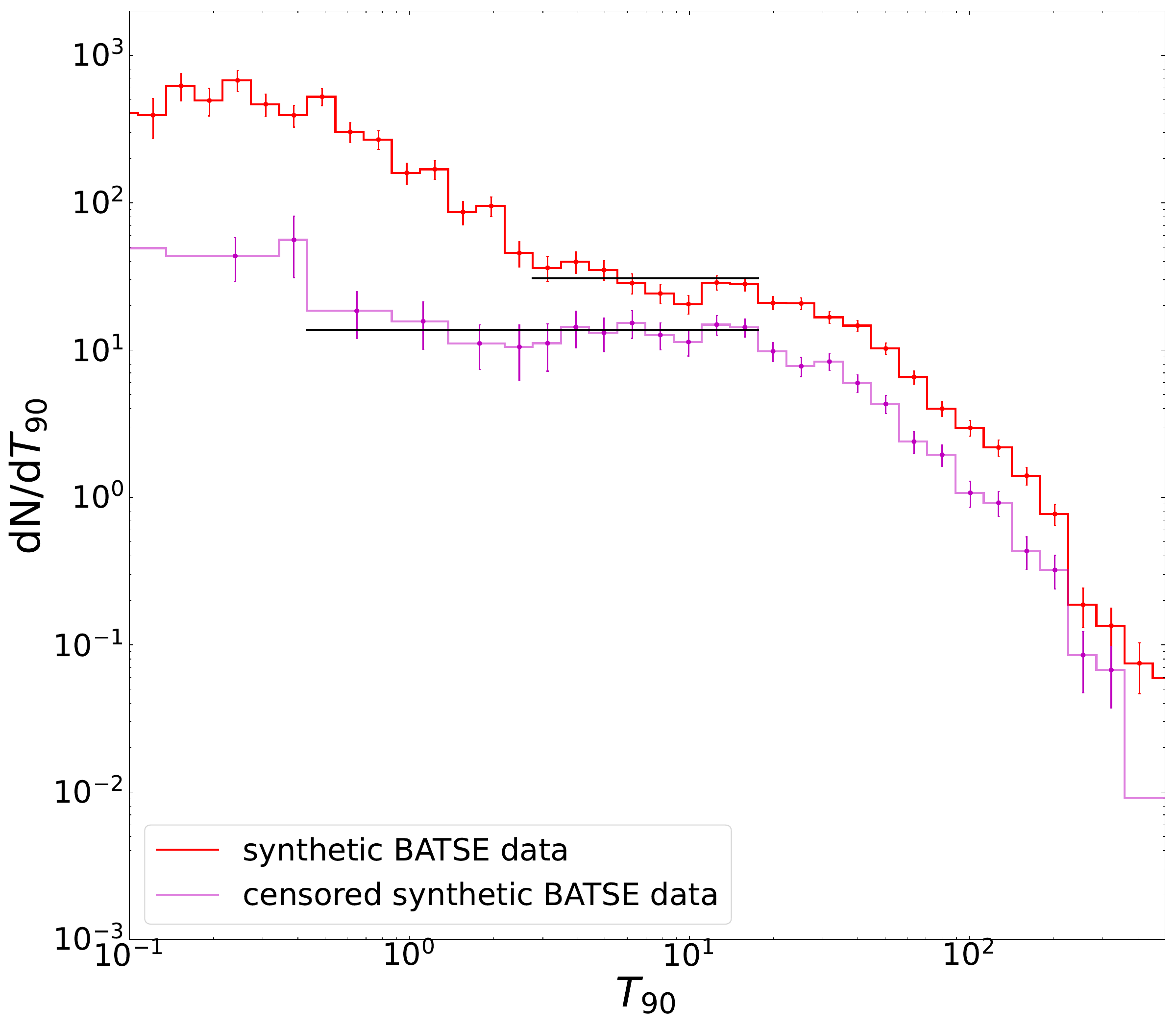}
		\caption{
			An illustration of the impact of excluding SGRBs from the duration distribution on the observed plateau of LGRBs. The original plateau in the red histogram and the extension of it by removing SGRBs in the magenta histogram are purely statistical (resulting from the Lognormal fits to data as depicted in Figure \ref{fig:allDNDTObserved}) and have no connections to the Physics of GRBs or Collapsars whatsoever.
			\label{fig:SyntheticSamplePlateau}
		}
	\end{figure}

	To resolve the sample contamination problem and minimize the impact of SGRBs duration distribution on the observed plateau in the duration distribution of LGRBs, \citet{bromberg2012observational} also restrict their analysis only to soft LGRBs in the BATSE catalog. This is done by removing any events with a hardness ratio \footnote{The hardness ratio is defined as the ratio of fluence between BATSE energy channels. In this case, channels 3 (100-300 keV) and 2 (50-100 keV)} are above $2.6$. This artificial cutoff preferentially excludes SGRBs from the duration histogram and significantly extends the observed plateau of LGRBs to more than twice the original length. However, in contrast to the arguments of \citet{bromberg2012observational}, this plateau extension in the population of soft LGRBs has no connection with the Collapsar interpretation of the plateau. It is merely an artifact of further censorships and sample incompleteness caused by the exclusion of an arbitrarily chosen subset of data. To illustrate this phenomenon in effect, we follow \cite{shahmoradi2013multivariate,shahmoradi2015short,osborne2020multilevelapj} to fit the BATSE catalog data with a comprehensive model for the duration, spectral peak energy, peak luminosity, and isotropic energy of GRBs. Notably, we assume that the duration distributions of both LGRBs and SGRBs follow Lognormal, making no assumption on the existence of plateaus in the duration distributions. Once the fitting is performed, we follow the prescription of \citet{bromberg2012observational} to remove the observed events from our Monte Carlo Universe with hardness ratios larger than $2.6$. Figure \ref{fig:SyntheticSamplePlateau} compares the resulting duration histogram for the soft population of LGRBs with the original histogram from the Monte Carlo Universe. A few remarks are in order:
	
	Firstly, plateaus appear in the duration distributions of both LGRBs and SGRBs without even requiring their existence in the model. Secondly, the LGRB plateau extends to about two orders of magnitude when we follow the prescription of \citet{bromberg2012observational} to exclude short hard bursts from the histogram. While purely statistical, this plateau extension is on par with the reported extension \citet{bromberg2012observational} found in the BATSE catalog data.
	
	The specific binning approach used to construct histograms of data also appears to moderately to significantly impact the strength and extent of any plateau in the short tails of distributions. This binning effect is well illustrated in Figures \ref{fig:lognormalB}, \ref{fig:weibullB}, and \ref{fig:fB}. The above arguments point to a non-physical origin for the observed plateau in the duration distribution of LGRBs. In fact, the distributions of other observational properties of GRBs (e.g., the observed redshift distribution as seen in Figure \ref{fig:zDnDz}) also exhibit plateaus without any apparent physical origins.
	
	The plateau emerging within the LGRB duration distribution is not unique and can also be seen in the population of SGRBs. Unlike the population of LGRBs, the decline in the short tail of the duration distribution of SGRBs is readily seen. This decline in the short tail actually suggests why we don't see such a fall in the short tails of LGRBs duration distribution. Unlike the case for LGRBs, no additional mixing distribution ``covers" this decline in the SGRB population. Such a plateau in the SGRB duration distribution is hard to reconcile with the non-collapsar origin of SGRBs. \citet{moharana2017observational} propose a similar prompt emission mechanism to that of LGRBs (by requiring the SGRB jet to drill through an envelope) to explain the non-collapsar plateau seen in the duration distribution of SGRBs. The likelihood ratio test used in \citet{moharana2017observational} for comparing the performances of the Lognormal and Plateau models to observational data only applies to nested statistical models, while the models compared are non-nested with entirely different parameters. To compare the goodness of fit of different models to the combined SGRB and LGRB BATSE data, we also consider four modeling scenarios: the Lognormal-Lognormal mixture model as illustrated in Figure \ref{fig:observedGRB}, the proposed plateau/powerlaw-plateau/powerlaw(tapered) mixture model as in Eqn. (3) in \citet{moharana2017observational}, Lognormal-plateau/powerlaw(tapered) mixture model, and plateau/powerlaw-Lognormal mixture model. The resulting fits yield the corresponding Bayesian Information Criterion scores: $8339$, $8354$, $8360$, $8362$, implying that the observational SGRB/LGRB data is $\sim3\times10^{6} - 10^{9}$ times more likely to have originated from a Lognormal-Lognormal mixture model than any alternative models with plateaus for the duration distributions of either or both GRB classes.
	
	By contrast, the statistical arguments presented in this manuscript offer a natural explanation for plateaus in the duration distributions of both GRB populations without recourse to any physical arguments and origins of GRBs. The fact that a wide variety of statistical distributions fit the duration distributions of GRBs equally well, as we demonstrated in this manuscript, further corroborates the findings of \cite{ghirlanda2015short, salafia2020gamma} and serves as reminders to exercise caution in attributing the observed plateaus in the duration distributions of GRBs to their physics.
	
	In sum, we have presented an alternative equally plausible purely statistical origin for the observed plateaus in the duration distributions of LGRBs and SGRBs. Our analysis further signifies the relevance and importance of selection effects and data censorship \citep[e.g.,][]{petrosian1996fluence, lloyd1999distribution, petrosian1999cosmological, lloyd2000cosmological, hakkila2003sample, band2005testing, nakar2004outliers, butler2007complete, shahmoradi2009real, butler2009generalized, butler2010cosmic, shahmoradi2011cosmological, shahmoradi2011possible, 2013arXiv1308.1097S, petrosian2015cosmological, coward2015selection, osborne2020multilevelapj, osborne2021there, bryant2021unbiased, tarnopolski2021does} in data-driven studies of the Physics of GRBs.

    \begin{acknowledgements}
        All authors contributed equally to this manuscript. FB acknowledges the support of the US National Science Foundation (NSF) under Grant No. 2138122. We appreciate the anonymous referee's careful and constructive comments, which significantly improved the quality of this manuscript's scientific contents.
    \end{acknowledgements}

    \bibliographystyle{aa}


\end{document}